# Chiral photonic crystals from sphere packing

Tao Liu,[1] Ho-Kei Chan (陈浩基),[2] and Duanduan Wan[1,*]

[1] School of Physics and Technology, Wuhan University, Wuhan 430072, China

[2] School of Science, Harbin Institute of Technology (Shenzhen), Shenzhen 518055, China

Corresponding Email [*]: ddwan@whu.edu.cn

**Abstract:** Inspired by recent developments in self-assembled chiral nanostructures, we have explored the possibility of using spherical particles packed in cylinders as building blocks for chiral photonic crystals. In particular, we focused on an array of parallel cylinders arranged in a perfect triangular lattice, each containing an identical densest sphere packing structure. Despite the non-chirality of both the spheres and cylinders, the self-assembled system can exhibit chirality due to spontaneous symmetry breaking during the assembly process. We have investigated the circular dichroism effects of the system and have found that, for both perfect electric conductor and dielectric spheres, the system can display dual-polarization photonic band gaps for circularly polarized light at normal incidence along the axis of the helix. Further, we have examined how the polarization band gap size depends on the dielectric constant of the spheres and the packing fraction of the cylinders. Our study suggests that a cluster formed by spheres self-assembling inside parallel cylinders with appropriate material parameters can be a promising approach to creating chiral photonic crystals.

The densest-packed structures of identical spheres in unbounded space include the well-known face-centered cubic (FCC) and hexagonal close-packed (HCP) structures. However, when cylindrical confinement is present, the densest-packed structures are in most cases helical, with the configuration depending on the ratio of $D/d$, where $D$ is the cylinder diameter and $d$ is the sphere diameter. Self-assembly of particles in a cylinder, forming helical crystalline structures, has been observed in a variety of systems, such as biological microstructures [1], fullerenes in nanotubes [2,3], nanoparticles in cylindrical confinements [4-6], foams in tubes [7,8], and colloidal particles or beads in microchannels [9-17].

Chirality, on the other hand, plays a crucial role in the interactions of materials with circularly polarized light. While chiral photonic crystals made from top-down, man-made helical structures exhibit polarization band gaps and have been widely explored [18-22], creating chiral nanomaterials from self-assembly schemes has become a subject of much interest. For example, a recent study [23] synthesized photonically active bowtie nanoassemblies with chirality continuum. Self-assembly approaches have advantages such as scalability, tunability, and cost-effectiveness (e.g., Refs. [23-28]).

Given the prevalence of self-assembly of particles in cylindrical geometries yielding chiral structures, can these structures be used to make chiral photonic crystals? To pursue an answer to this question, here we investigated the feasibility of using this bottom-up approach to create chiral photonic crystals through computer simulations. We considered an array of parallel cylinders arranged in a perfect triangular lattice, each containing an identical densest sphere packing structure. We have investigated the circular dichroism (CD) effects of the system and find that for both perfect electric conductor (PEC) and dielectric spheres, the system can display dual-polarization photonic band gaps for circularly polarized light at normal incidence along the helical axis direction. We have further looked into the dependence of the polarization band gap size on the dielectric constant of the spheres and the packing fraction of the cylinders.

It is well known that entropy plays an important role in the formation of phases in colloidal dispersions [29]. When geometrical confinement is present, the self-assembled structures can be different from those in bulk [29-34]. In the case of the self-assembly of spheres in the presence of a cylinder, despite the non-chirality of both the spheres and cylinders, the self-assembled structure of spheres in a cylinder can exhibit chirality due to spontaneous symmetry breaking during the assembly process (e.g., Refs. [11-17]). Here we have considered the densest-packed structures of monodisperse hard spheres in a cylinder. These structures have been previously identified, which depend on the ratio of the cylinder diameter $D$ to the hard-sphere diameter $d$ [35-41]. The configurations of dipolar hard spheres can be found in Ref. [42]. In Table I, we list the first eight densest structures described in Ref. [37], corresponding to the ratio $1 \leq D/d \leq 2.141$. Among them, structures 3, 4, 6, 7, and 8 are chiral. Because the longitudinal lattice constant of structure 7 is very long, making it computationally expensive, we only focus on the chiral structures 3, 4, 6, and 8. We take $D/d$ =1.893, 1.996, 2.038, and 2.044, and use 9, 26, 30, and 30 spheres as a primitive cell, for the four structures, respectively [Fig. 1(a-d)]. With these choices, the total helix angle for the spheres in a primitive cell is a multiple of $2\pi$ (see the Supplementary Material for more details).

We then considered a situation where cylinders are laterally arranged in a perfect triangular lattice, forming a crystal [Fig. 1(e) and 1(f)]. To realize an array of parallel cylinders, one approach is to extract micropores in a polymer film [15,16]. Another way is to pack microtubes. When tubes are long enough and at high packing densities, they tend to align parallel to each other [29,30,43-45], and their cross sections form a triangular lattice. Experiments have obtained monodomain films of highly aligned carbon nanotubes from suspension [46]. We expect that, in a similar manner, by taking advantage of entropic effects and using auxiliary experimental skills, it is possible to align long microtubes as well. For simplicity, here we assume that the cylinders have a negligible thickness and possess identical sphere packing structures inside. We take the triangular lattice constant $a = 1.1 \times D$ as an example. This choice results in a packing fraction for the cylinders $\phi = \pi D^2 / 2\sqrt{3}a^2 \approx 0.75$. Note this is not the packing fraction of spheres. Unless otherwise specified, the circular polarization properties of the Bloch modes are analyzed by full-field numerical simulations using the software COMSOL Multiphysics.

Figure 2(a) plots the crystal's transmission spectra and band structure that consists of structure 3 in the triangular lattice with PEC spheres inside. The color and size in the band structure represent the CD index $\alpha$ and the coupling index $\beta$, as defined in previous studies [22,47] (see Supplementary Material for more information). The value of $\alpha$ falls within the range of [-1, 1], where $\alpha = +1$ represents a pure right-handed (RH) circularly polarized mode and $\alpha = -1$ represents a pure left-handed (LH) circularly polarized mode. $\beta$ evaluates the efficiency of coupling a plane wave into the Bloch mode. The value of $\beta$ ranges from 0 to 1, where a small value indicates a low coupling efficiency of the mode. Because the CD of the two chiral structures, i.e., LH or RH, is dual, we only considered either of the two dual chiral structures, but not both. In practice, the preferred chirality can be controlled through non-equilibrium self-assembly in rotating fluids [17]. The band structure along the $\Gamma$-A direction, i.e., along the axis of the helix [Fig. 1(g)], was calculated. It can be seen that there is a LH gap at the edge of the first Brillouin zone ($\Omega$=0.161 to 0.293) and a RH gap at the edge of the second Brillouin zone ($\Omega$=0.408 to 0.438). No complete polarization gap for both LH and RH gaps is observed. Figure 2(b) plots the case of structure 4. As shown in the figure, there are two LH gaps, i.e., $\Omega$=0.521 to 0.716 $\Omega$=0.755 to 0.828. Besides, we observe a wide RH gap at $\Omega$=0.404 to 0.667. Thus, this gives rise to a complete polarization gap ($\Omega$=0.521 to 0.667). Similarly, for structure 6, there is a single LH gap ($\Omega$=0.573 to 0.695) and two RH gaps ($\Omega$=0.307 to 0.399 and $\Omega$=0.664 to 0.695), which yields a complete polarization gap at $\Omega$=0.664 to 0.695 [Fig. 2(c)]. For structure 8, as shown in Fig. 2(d), there is a wide LH gap at $\Omega$=1.102 to 1.257, and two RH gaps at $\Omega$=0.701 to 0.879 and $\Omega$=1.177 to 1.257, respectively. This gives a complete polarization gap at $\Omega$=1.177 to 1.257. The polarization gaps for PEC spheres are summarized in Table II.

We have also explored a scenario where the spheres are made of dielectric materials. Figure 3 shows parallel results as those in Fig. 2, but with dielectric spheres inside the cylinders. We have optimized the dielectric constant to maximize the first polarization band gap size. The optimal dielectric constants for structures 3, 4, 6, and 8 are 24, 6, 44, and 44, respectively. For the crystal corresponding to structure 3 [Fig. 3(a)], we have observed a LH gap at a low-frequency range $\Omega$=0.147 to 0.171 and a RH gap at a higher frequency range $\Omega$=0.283 to 0.289. It is worth noting that although the band structure of photonic crystals made of dielectric spheres at low frequencies is similar to that of PEC spheres for this structure, the size of the polarization band gap is much smaller. In the case of structure 4, as shown in Fig. 3(b), the first polarization gap is a RH gap at $\Omega$=0.462 to 0.477, and the second polarization gap is a LH gap at $\Omega$=0.529 to 0.551. The fact that the optimal dielectric constant is 6 suggests that materials with low dielectric constants, such as silica [11,14,16] and polymeric materials [9,10,12,14,15], can achieve a relatively large polarization bandgap. For structure 6, as shown in Fig. 3(c), the first polarization gap is a LH gap at $\Omega$=0.282 to 0.287, and the second polarization gap is a RH gap at 0.294 to 0.296. In the case of structure 8 [Fig. 3(d)], there is a LH gap at $\Omega$=0.284 to 0.288 and a RH gap at $\Omega$=0.295 to 0.297. It is worth noting that structure 6 and structure 8 share the same optimal dielectric constant value. Moreover, a comparison of Fig. 3(c) and 3(d) reveals that they have similar band structures, and the frequencies at which the polarization gaps appear are close. We expect that this similarity may be due to the similarity of their structures. The polarization gaps for dielectric spheres are summarized in Table III.

To investigate how the size of the polarization gap depends on the dielectric constant and the packing fraction of the cylinders, we take structure 3 as an example. We have calculated the relative gap size (defined as $\Delta\Omega/\Omega_0$, where $\Delta\Omega$ is the width of the polarization bandgap and $\Omega_0$ is the central frequency) as a function of the packing fraction $\phi$ and the dielectric constant $\varepsilon$. We have explored a wide range of $\phi$ and $\varepsilon$, with $\phi$ ranging from 0.15 to 0.9, and $\varepsilon$ from 2 to 60. To accelerate the computation, we employed the open-source software MIT PHOTONIC BANDS [48] to compute the band structure.

As shown in Fig. 4, for a given $\varepsilon$ value, the largest relative gap size occurs at the maximal packing fraction $\phi$ = 0.9. Besides, there is a general trend that the relative gap size increases as the packing fraction increases, except for small to medium packing fraction regions where the relative gap size remains relatively small (i.e., smaller than or approximately 7%). Furthermore, for relatively large packing fractions (i.e., $\phi \geq 0.7$), the relative gap size first increases and then decreases as the dielectric constant increases. The largest relative gap size observed in the diagram is approximately 20%, which suggests the potential of achieving a large polarization band gap by properly choosing the $\phi$ and $\varepsilon$ values.

In conclusion, inspired by the self-assembled helical structures of spheres in a cylinder, we have explored the possibility of using these structures as building blocks for chiral photonic crystals. Specifically, four kinds of building blocks were considered, being the densest packing structures of hard spheres within a cylinder. We have examined CD effects in photonic crystals composed of these building blocks arranged in a two-dimensional triangular lattice, and we have found that for both PEC and dielectric spheres, the system can exhibit dual-polarization photonic band gaps for circularly polarized light at normal incidence along the helical axis direction. Moreover, we have explored the dependence of the polarization band gap size on the dielectric constant of the spheres and the packing fraction of the cylinders, and have found that large polarization band gaps can be achieved. This proposed system has several advantages, such as it is relatively easy to make spherical particles in experiments, the configuration of the building blocks can be adjusted by designing the $D/d$ value, and the particle size can be varied to tune the periodicity of crystals and consequently the photonic band gap frequency. In all, our study suggests that the self-assembly of spherical particles in cylindrical channels with appropriate material parameters can be a promising approach to creating chiral photonic crystals.

# Acknowledgements

This work was supported by the National Natural Science Foundation of China (Grant No. 12274330, 11904265), and the Knowledge Innovation Program of Wuhan-Shuguang (Grant No. 2022010801020125).

# Tables

**Table I.** The first eight densest-packed structures of hard spheres in a cylinder from Ref. [37].

| Structure | Range | Average contact number | Chirality |
|---|---|---|---|
| 1 | $D/d = 1$ | 2 | Achiral |
| 2 | $1 < D/d \leq 1.866$ | 2 | Achiral |
| 3 | $1.866 < D/d \leq 1.995$ | 4 | Chiral |
| 4 | $1.995 \leq D/d < 2.0$ | 4 | Chiral |
| 5 | $D/d = 2.0$ | 5 | Achiral |
| 6 | $2.0 < D/d < 2.039$ | 5 | Chiral |
| 7 | $D/d = 2.039$ | 6 | Chiral |
| 8 | $2.039 < D/d \leq 2.141$ | 5 | Chiral |

**Table II.** Polarization gaps for PEC spheres. Numbers in brackets are the relative gap size $\Delta\Omega/\Omega_0$.

| PEC spheres | LH gap | RH gap | complete polarization gap |
|---|---|---|---|
| Structure 3 | $\Omega = 0.161 \sim 0.293$ (58.15%) | $\Omega = 0.408 \sim 0.438$ (7.09%) | ----- |
| Structure 4 | $\Omega = 0.521 \sim 0.716$ (31.53%)<br>$\Omega = 0.755 \sim 0.828$ (9.22%) | $\Omega = 0.404 \sim 0.667$ (49.11%) | $\Omega = 0.521 \sim 0.667$ (24.58%) |
| Structure 6 | $\Omega = 0.573 \sim 0.695$ (19.24%) | $\Omega = 0.307 \sim 0.399$ (26.06%)<br>$\Omega = 0.664 \sim 0.695$ (4.56%) | $\Omega = 0.664 \sim 0.695$ (4.56%) |
| Structure 8 | $\Omega = 1.102 \sim 1.257$ (13.14%) | $\Omega = 0.701 \sim 0.879$ (22.53%)<br>$\Omega = 1.177 \sim 1.257$ (6.57%) | $\Omega = 1.177 \sim 1.257$ (6.57%) |

**Table III.** Polarization gaps for dielectric spheres. Numbers in brackets are the relative gap size $\Delta\Omega/\Omega_0$.

| dielectric spheres | LH gap | RH gap | complete polarization gap |
|---|---|---|---|
| Structure 3 | $\Omega = 0.147 \sim 0.171$ (15.09%) | $\Omega = 0.283 \sim 0.289$ (2.10%) | ----- |
| Structure 4 | $\Omega = 0.529 \sim 0.551$ (4.07%) | $\Omega = 0.462 \sim 0.477$ (3.19%) | ----- |
| Structure 6 | $\Omega = 0.282 \sim 0.287$ (1.76%) | $\Omega = 0.294 \sim 0.296$ (0.68%) | ----- |
| Structure 8 | $\Omega = 0.284 \sim 0.288$ (1.40%) | $\Omega = 0.295 \sim 0.297$ (0.68%) | ----- |

# Figures

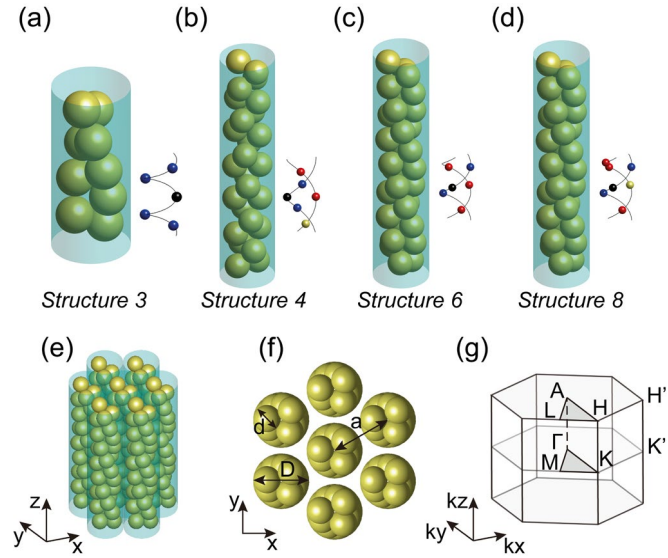

FIG. 1 (Color online) (a-d) Four kinds of chiral structures from the densest-packed structures of identical hard spheres in a cylinder studied in this work. Each figure shows a primitive cell, corresponding to structures 3, 4, 6, and 8 in Table I, respectively. The spheres are drawn in yellow and the cylinder in cyan. The insets show the helices of the structures (black lines) and local conditions of a particle (marked in black), where particles in blue and red are in contact with the black particle, but on the same and the other spiral, respectively. Particles in yellow are not contacting the black particle. (e) An array of cylinders laterally arranged in a perfect triangular lattice. (f) Top view of the array. (g) The first Brillouin zone. $\Gamma$ and $A$ are high-symmetry points along the helical axis direction.

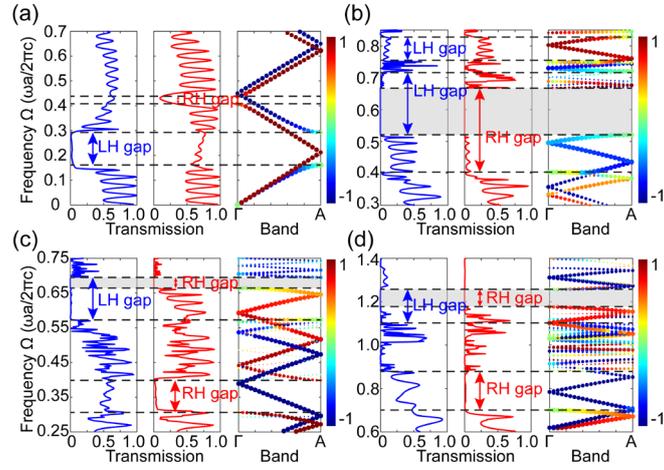

FIG. 2 (Color online) Transmission spectra and band structures along the Γ-A direction of the crystal that consists of structure 3 (a), structure 4 (b), structure 6 (c), and structure 8 (d) in the triangular lattice with PEC spheres inside, respectively. For transmission spectra calculation, LH (blue) and RH (red) circularly polarized light propagates from bottom to top, normally along the helical axis direction. For band structures, the color and size of the symbols indicate the CD index $\alpha$ and the coupling index $\beta$, respectively.

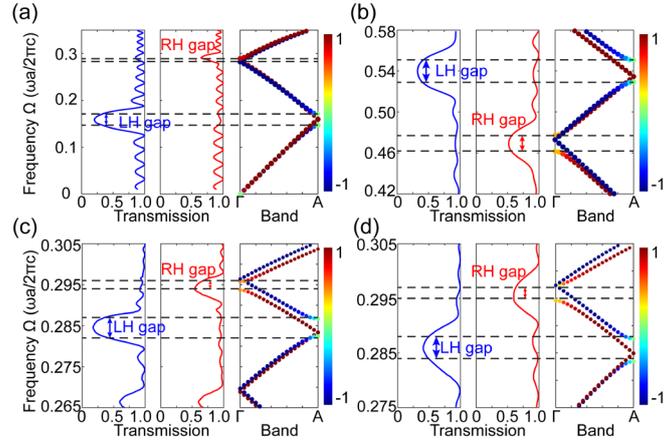

FIG. 3 (Color online) Transmission spectra and band structures along the Γ-A direction of the crystal that consists of structure 3 (a), structure 4 (b), structure 6 (c), and structure 8 (d) in the triangular lattice with dielectric spheres inside, respectively. The relative dielectric constant in (a-d) is $\varepsilon$ = 24, 6, 44 and 44, respectively.

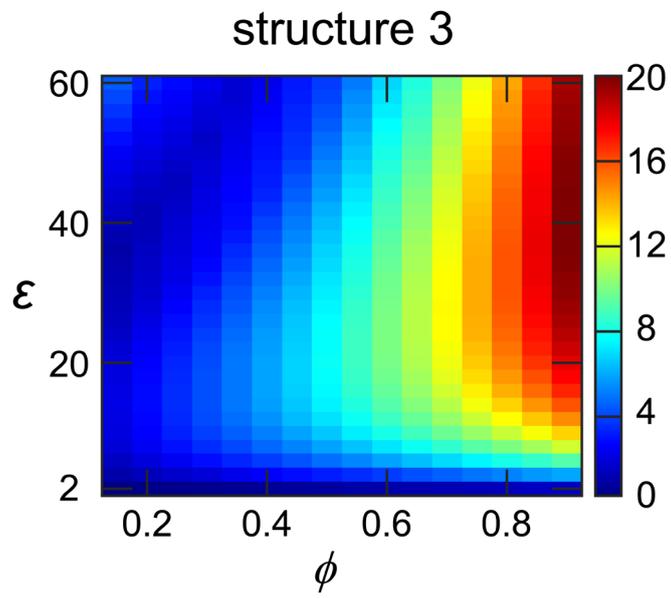

FIG. 4 (Color online) The relative gap size of the first polarization gap as a function of $\phi$ and dielectric $\varepsilon$ for structure 3.

# Supplementary Material for: Chiral photonic crystals from sphere packing

Tao Liu,[1] Ho-Kei Chan (陈浩基),[2] and Duanduan Wan[1,*]

[1] School of Physics and Technology, Wuhan University, Wuhan 430072, China

[2] School of Science, Harbin Institute of Technology (Shenzhen), Shenzhen 518055, China

Corresponding Email [*]: ddwan@whu.edu.cn

## I. The densest-packed structures of identical hard spheres in a cylinder

Figure 1 in the main text displays the four chiral structures investigated in this study.

(1) Structure 3 is the first chiral structure to occur when the ratio of cylinder diameter to sphere diameter $D/d$ increases from 1. Its $D/d$ range is $1.866 < D/d \leq 1.995$. Each sphere in structure 3 is in contact with two spheres above and two spheres below on a single helix. The average number of contacts is 4. Structure 3 can also be seen as two helices coiling around each other [1]. Here we consider it as a single helix and use $D/d = 1.893$ for calculation. The structure then consists of spheres arranged on a helix, and each primitive cell contains 9 spheres. The total helix angle for the spheres in a primitive cell is $4 \times 2\pi$.

(2) Structure 4 corresponds to the range $1.995 \leq D/d < 2.0$. Each sphere is in contact with a sphere above and a sphere below on the same helix, as well as two spheres on the other helix. The average number of contacts is 4. We use $D/d = 1.996$ for calculation. The structure then is composed of two identical staggered helices, and each primitive cell contains 26 spheres. The total helix angle for the spheres in a primitive cell is $3 \times 2\pi$.

(3) Structures 6 and 8 correspond to the ranges $2.0 < D/d < 2.039$ and $2.039 < D/d \leq 2.141$, respectively. These two structures can be regarded as two staggered helices, resulting from the line-slip deformation of a single helix (structure 7) [2]. In both structures, each sphere is in contact with a sphere above and a sphere below on the same helix, and three spheres on the other helix. The average number of contacts is 5. The difference is that for structure 6, the three adjacent spheres on the other helix are arranged in sequence, while for structure 8, only two out of the three spheres on the other helix are in contact with each other. We use $D/d = 2.038$ for structure 6 and $D/d = 2.044$ for structure 8. Both structures contain 30 spheres in each primitive cell, and the total helix angle for the spheres in a primitive cell is $4 \times 2\pi$.

## II. Circular polarization analysis and transmission calculation

The normalized circularly polarized coupling coefficient is defined as [3,4]

$$\eta_{RH(LH)} = \frac{|\iint [\frac{1}{\sqrt{2}}(\hat{a}_x \mp i\hat{a}_y)] \cdot \vec{H}(x,y,z_0) \, dxdy|^2}{\iint |\frac{1}{\sqrt{2}}(\hat{a}_x \mp i\hat{a}_y)|^2 \, dxdy \iint |\vec{H}(x,y,z_0)|^2 \, dxdy}, \qquad (1)$$

which represents the overlap integrals between the magnetic field $\vec{H}(x, y, z_0)$ of Bloch modes at $z = z_0$ plane and the circularly polarized (CP) waves, described by right-handed (RH) circularly polarized light $(\hat{a}_x - i\hat{a}_y)/\sqrt{2}$ or left-handed (LH) circularly polarized light $(\hat{a}_x + i\hat{a}_y)/\sqrt{2}$.

The circular dichroism (CD) index is

$$\alpha = \text{sgn}(\vec{q} \cdot \nabla_k \omega) \frac{\eta_{RH} - \eta_{LH}}{\eta_{RH} + \eta_{LH}}, \quad (2)$$

where sgn is the sign function and $\vec{q}$ is the wave vector of the incident circularly polarized wave. With this definition, the value of $\alpha$ falls within the range of [-1, 1], where $\alpha = +1$ represents a pure RH circularly polarized mode and $\alpha = -1$ represents a pure LH circularly polarized mode.

The efficiency of coupling a plane wave into the Bloch mode is evaluated using the coupling index

$$\beta = \eta_{RH} + \eta_{LH}. \quad (3)$$

The value of $\beta$ ranges from 0 to 1, where a small value indicates a low coupling efficiency of the mode.

To obtain the transmission spectra, the Floquet periodic boundary conditions are applied around the structure, and perfectly matched layers (PML) are added to the top and bottom surfaces to absorb residual reflected and transmitted light. The RH and LH circularly polarized light are normal incident from bottom to top along the z direction, respectively. The corresponding transmittance is calculated as the ratio of the intensity of outgoing RH/LH light to incident RH/LH light, $T^{RH(LH)} = I_{out}^{RH(LH)} / I_{in}^{RH(LH)}$, obtained from the scattering parameters. In practice, for structure 3, five unit cells along the z direction are used for the calculation, while for structures 4, 6, and 8, two unit cells are used. This is to ensure that the polarization band gap is recognized.